\documentclass[aps,prl,reprint,superscriptaddress,showpacs]{revtex4-1}
\usepackage{graphicx,amsmath,amssymb,ulem}
\usepackage{verbatim}
\usepackage{ragged2e}
\usepackage{color}

\begin{document}

\title{Schmidt modes in the angular spectrum of bright squeezed vacuum}
\author{P.~Sharapova}
\affiliation{Physics Department, Moscow State University, Leninskiye Gory 1-2, Moscow 119991, Russia}
\author{A.~M.~P\'erez}
\affiliation{Max-Planck Institute for the Science of Light, \\ Guenther-Scharowsky-Str. 1 / Bau 24, Erlangen  D-91058, Germany}
\author{O.~V.~Tikhonova}
\affiliation{Physics Department, Moscow State University,  Leninskiye Gory 1-2, Moscow 119991, Russia}
\affiliation{Skobeltsyn Institute of Nuclear Physics, Lomonosov Moscow State University, Moscow 119234, Russia}
\author{M.~V.~Chekhova}
\affiliation{Max-Planck Institute for the Science of Light, \\  Guenther-Scharowsky-Str. 1 / Bau 24, Erlangen  D-91058, Germany}
\affiliation{Physics Department, Moscow State University, Leninskiye Gory 1-2, Moscow 119991, Russia}

\begin{abstract}
 We study the spatial properties of high-gain parametric down-conversion (PDC). From the Hamiltonian we find the Schmidt modes, apply the Bloch-Messiah reduction, and calculate analytically the measurable quantities, such as the angular distributions of photon numbers and photon-number correlations. Our approach shows that the Schmidt modes of PDC radiation can be considered the same as for the low-gain (biphoton) case while the Schmidt eigenvalues strongly depend on the parametric gain. Its validity is confirmed by comparison with several experimental results, obtained by us and by other groups.
\end{abstract}
\maketitle

\textit{Introduction.} Currently, much interest is attracted to bright squeezed vacuum (BSV), a macroscopic non-classical state of light that can be obtained via high-gain parametric down conversion (PDC). This is because BSV contains huge photon numbers and at the same time is strongly nonclassical, manifesting entanglement~\cite{nonsep} and noise reduction below the standard quantum limit~\cite{two-color}. These features make it interesting for applications such as quantum imaging~\cite{QI,Q-illum} and metrology~\cite{metro}, quantum optomechanics~\cite{optomech} and nonlinear optics with quantum light~\cite{diTrapani}. Besides, large photon-number correlations arising in BSV are richer than entanglement in two-photon light  emitted via low-gain PDC~\cite{richer}.

At the same time, theoretical description of BSV presents certain difficulties. In contrast to low-gain PDC, BSV generation cannot be described in the framework of the perturbation theory. The quantum state contains not only two-photon terms but also higher-order Fock components, and its calculation in the Schr\"odinger picture is difficult.
Many recent theoretical investigations of such strong pumping regime are based on the concept of collective input/output modes introduced mostly to describe the spectral properties in the frequency domain~\cite{Boyd,Wasilewski,Dayan,Christ,Eckstein}. In a high-gain regime it is convenient to calculate the observables in the Heisenberg picture. In this case the Schmidt-mode formalism used in the Schroedinger picture for the description of multimode two-photon light~\cite{Eberly} is  replaced by a similar procedure, called Bloch-Messiah reduction~\cite{Braunstein}.  However beyond the perturbation approach the solution for the field operators was up to now obtained only numerically, usually through a set of integro-differential equations~\cite{Christ,Eckstein,Brambilla2004,Brambilla2008,Brambilla2010}.

Here we present a fully analytical description of the angular properties of BSV. Our approach is based on the Bloch-Messiah reduction and allows one to obtain the evolution of the photon creation (annihilation) operators both for the Schmidt modes and for the plane-wave modes. After obtaining the evolution of these, we calculate analytically the angular distributions of the intensity and various correlation characteristics of the down-converted radiation. It should be emphasized that the applied formalism allows one to find the effective number of independent collective modes of the system, a measure of its information capacity. The  approach is applied to the description of several observable effects: photon-number correlations in the angular spectrum, their dependence on the parametric gain, and the manifestation of the transverse walk-off.

\textit{Bloch-Messiah reduction.} At any parametric gain, PDC is described by the Hamiltonian of the form
\begin{equation}
H\sim\int d^3\mathbf{r}\chi^{(2)}(\mathbf{r})E_p^{(+)}(\mathbf{r},t)E_s^{(-)}(\mathbf{r},t)E_i^{(-)}(\mathbf{r},t)+h.c.,\nonumber
\label{Ham}
\end{equation}
where $\chi^{(2)}(\mathbf{r})$ is the quadratic nonlinearity and $s,i,p$ indices  correspond to the signal, idler, and pump radiation, respectively. After integrating over space, assuming that the pump amplitude has a Gaussian spatial distribution, we obtain  the Hamiltonian in the simplest case of type-I PDC in one non-anisotropic nonlinear crystal:
\begin{equation}
H=i\hbar\Gamma\int d \mathbf{q}_s  d \mathbf{q}_i F(\mathbf{q}_s,\mathbf{q}_i)a^\dagger_{\mathbf{q}_s}a^\dagger_{\mathbf{q}_i}+h.c.,
\label{Ham1}
\end{equation}
with the two-photon amplitude (TPA)
\begin{eqnarray}
F(\mathbf{q}_s,\mathbf{q}_i)=C \exp\{-\frac{\sigma^2 (\mathbf{q}_s+\mathbf{q}_i)^2}{2}\}\mathrm{sinc}(\frac{L(\mathbf{q}_s-\mathbf{q}_i)^2}{4 k_p})\times
\nonumber\\
\times\exp\{i\frac{ L(\mathbf{q}_s-\mathbf{q}_i)^2}{4 k_p} \}, \ \ \ \ \ \
\label{TPA}
\end{eqnarray}
where $\Gamma$ is the coupling constant, $\mathbf{q}_{s,i}$ the transverse components of the wave vectors  for the signal and idler beams, $a^\dagger_{\mathbf{q}_{s,i}}$ the photon creation operators for the corresponding plane-wave modes, $C$ the normalization constant, $2\sqrt{ln 2}\sigma$ the full width at half maximum (FWHM) of the pump spatial intensity distribution, $L$ the crystal length, and $k_p$ the pump wave vector.

For simplicity we assume here that $\omega_s=\omega_i=\dfrac{\omega_p}{2}$, so that the TPA depends only on the transverse wave vectors of the signal and idler beams. Next, we apply the Schmidt decomposition to $F(\mathbf{q}_s,\mathbf{q}_i)$, which is not simple if the full dimensionality is taken into account. The Schmidt decomposition can be made explicitly in the cylindrical frame of reference or by using the double-Gauss approximation~\cite{Law,Fedorov,double-Gauss} in the Cartesian frame. In the latter case the \textit{sinc} factor in (\ref{TPA}) is approximated by a Gaussian function and then the TPA can be factorized in two terms, each one depending only on $x$ or $y$ components of $\mathbf{q}_s,\mathbf{q}_i$. Using analytical Schmidt decomposition in each direction, we obtain the total decomposition of the TPA $F(\mathbf{q}_s,\mathbf{q}_i)$~\cite{sup}. Here we will focus on the explicit Schmidt decomposition performed in the cylindrical frame of reference, in which each transverse wave vector is given by its absolute value $q_{i,s}$ and azimuthal angle $\phi_{i,s}$. In this case,
$F(\mathbf{q}_s,\mathbf{q}_i)$ can be written as a Fourier expansion due to its periodicity in $(\phi_s-\phi_i)$~\cite{Miatto},
\begin{equation}
F(q_s,q_i,\phi_s-\phi_i)=\sum_n \chi_n(q_s,q_i) e^{i n (\phi_s-\phi_i)},
\label{TPA polar}
\end{equation}
where $\chi_n(q_s,q_i)$ can be found using the inverse Fourier transformation. Then, the 1-dimensional Schmidt decomposition yields
\begin{equation}
\chi_n(q_s,q_i)=\sum_m \sqrt{\lambda_{mn}}\frac{u_{mn}(q_s)}{\sqrt{q_s}} \frac{v_{mn}(q_i)}{\sqrt{q_i}},
\label{decomposition}
\end{equation}
with the functions $u_{mn}(q_s)$ and $v_{mn}(q_i)$ obeying the normalization condition,
\begin{equation}
\int_0^{\infty} d q_s u_{mn}(q_s)u^{*}_{kn}(q_s)=\int_0^{\infty} d q_i v_{mn}(q_i)v^{*}_{kn}(q_i)=\delta_{mk}.\nonumber
\label{orthogonalisation}
\end{equation}

Decomposition ~(\ref{decomposition}) allows one to expand the TPA as
\begin{equation}
F(\mathbf{q}_s,\mathbf{q}_i)=\sum_{m,n} \sqrt{\lambda_{mn}}\frac{u_{mn}(q_s)}{\sqrt{q_s}} \frac{v_{mn}(q_i)}{\sqrt{q_i}}e^{i n (\phi_s-\phi_i)}.
\label{Schmidt}
\end{equation}

Using Eq.~(\ref{Schmidt}), we rewrite the Hamiltonian ~(\ref{Ham1}) in terms of the photon creation/annihilation operators for the new collective spatial Schmidt modes,
\begin{equation}
H=i\hbar\Gamma \sum_{m,n} \sqrt{\lambda_{mn}}(A_{mn}^\dagger B_{mn}^\dagger-A_{mn} B_{mn}),
\label{Hams}
\end{equation}
where
\begin{eqnarray}
A_{mn}^\dagger=\int d\mathbf{q}_s \frac{u_{mn}(q_s)}{\sqrt{q_s}}e^{i n \phi_s}a^\dagger_{\mathbf{q}_s},
\nonumber\\
B_{mn}^\dagger=\int d\mathbf{q}_i \frac{v_{mn}(q_i)}{\sqrt{q_i}}e^{-i n \phi_i}a^\dagger_{\mathbf{q}_i},
\label{ops}
\end{eqnarray}
with $d\mathbf{q}=qd qd\phi$.

The performed procedure is the generalized Bloch-Messiah reduction, which in the frequency domain introduces new collective photon operators and collective modes called "broadband modes"  \cite{Christ,Eckstein}.

By passing to the new operators, we diagonalize the Hamiltonian~(\ref{Hams}), so that the modes corresponding to the new operators are coupled pairwise. Note that the new operators obey standard commutation relations:
\begin{equation}
[ A_{mn},A_{kl}^\dagger]=\delta_{mk} \delta_{nl},\,\,
[ A_{mn},B_{kl}^\dagger]=\delta_{mk} \delta_{n,-l}.
\label{comm}
\end{equation}
The last relation results from the non-distinguishability of the signal and idler photons, manifested in the invariance of amplitude ~(\ref{TPA}) to the $\mathbf{q}_s\longleftrightarrow  \mathbf{q}_i$ exchange. Further we will take this into account  by setting $u_{mn}(\xi)=v_{mn}(\xi)$.

The Heisenberg equations for the Schmidt-mode operators have the form
\begin{eqnarray}
\frac{d A_{mn}}{dt}=\Gamma \sqrt{\lambda_{m,n}}(A_{m,-n}^\dagger + B_{mn}^\dagger)=2\Gamma \sqrt{\lambda_{mn}}B_{mn}^\dagger , \ \
\nonumber\\
\frac{d B_{mn}^\dagger}{dt}=\Gamma \sqrt{\lambda_{m,n}}(A_{mn} + B_{m,-n})=2\Gamma \sqrt{\lambda_{mn}}A_{mn}. \ \ \
\label{Heis}
\end{eqnarray}
Here we used the relation $A_{m,-n}^\dagger = B_{mn}^\dagger$, $ A_{mn}=B_{m,-n}$, $\lambda_{mn}=\lambda_{m,-n}$.

The solutions are given by the Bogolyubov-type transformations,
\begin{eqnarray}
A_{mn}^{out}=A_{mn}^{in}\cosh[G\sqrt{\lambda_{mn}}]
+[B_{mn}^{in}]^\dagger \sinh[G\sqrt{\lambda_{mn}}], \nonumber\\
B_{mn}^{out}=B_{mn}^{in}\cosh[G\sqrt{\lambda_{mn}}]
+[A_{mn}^{in}]^\dagger \sinh[G\sqrt{\lambda_{mn}}], \nonumber
\label{Bogol}
\end{eqnarray}
where $G\equiv\int2\Gamma dt$ is the parametric gain and $A_{mn}^{in},B_{mn}^{in}$ are the initial (vacuum) photon annihilation operators in the corresponding Schmidt mode~(\ref{ops}).
Because in experiment one usually deals with plane-wave modes, visualized, for instance, in the focal plane of a lens, it is useful to obtain explicit expressions for the plane-wave photon creation and annihilation operators of the signal and idler beams. We do this analytically by expanding them over the set of orthogonal normalized Schmidt modes, and the explicit expressions for $a^\dagger_{\mathbf{q}_s}$, $a^\dagger_{\mathbf{q}_i}$ are given in the Supplemental Material. Further, we calculate all the observables using these time-dependent plane-wave operators.


Mean values of photon numbers and their second-order moments are found by averaging over the vacuum state,
\begin{equation}
\langle N_{s}(\mathbf{q}_s)\rangle= \langle 0 \vert a^\dagger_{\mathbf{q}_s} a_{\mathbf{q}_s} \vert 0 \rangle,
\label{mean}
\end{equation}
and similarly for $\langle N^{2}_{s}(\mathbf{q}_s)\rangle$ and $\langle N_{s}(\mathbf{q}_s) N_{i}(\mathbf{q}_i)\rangle$.

The obtained photon-number distributions for the signal and idler beams take the form
\begin{eqnarray}
\langle N_{s,i}(\mathbf{q}_{s,i})\rangle=\sum_{mn} \frac{\vert u_{mn}(q_{s,i})\vert^{2}}{q_{s,i}}(\sinh[G\sqrt{\lambda_{mn}}])^{2}.
\label{num}
\end{eqnarray}
This means that the Schmidt modes contribute to the total photon number incoherently, with the normalized weights
\begin{equation}
\tilde{\lambda}_{mn}=\frac{(\sinh[G\sqrt{\lambda_{mn}}])^2}{\sum_{mn}(\sinh[G\sqrt{\lambda_{mn}}])^2},
\label{new_Schmidt}
\end{equation}
which can be understood as the new Schmidt eigenvalues, renormalized at high gain. At low gain, $G\ll1$, $\tilde{\lambda}_n=\lambda_n$.

Finally, one can calculate the variance of the photon-number difference in the signal and idler beams, $\mathrm{Var}(N_{s}-N_{i})\equiv\langle (N_{s}-N_{i})^{2}\rangle-\langle N_{s}-N_{i}\rangle^{2}$, the correlation function $G_{is}^{(2)}\equiv\langle N_{s}N_{i}\rangle$, its normalized version and any other statistical values~\cite{sup}.

The expression for the TPA~(\ref{TPA}) can be generalized for more complicated experimental schemes. Our approach is valid whenever the amplitude $F$ is periodic in $(\phi_s-\phi_i)$.

Further, we apply our model to the description of three different experiments.

\begin{figure}[htb]
\begin{center}
\includegraphics[width=0.4\textwidth]{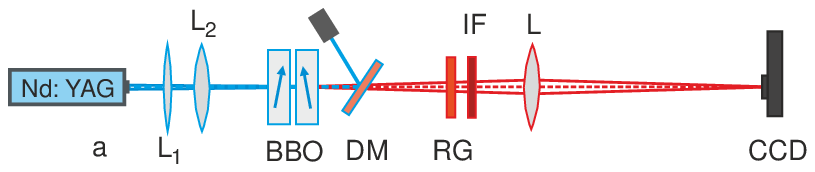}\\
\includegraphics[width=0.4\textwidth]{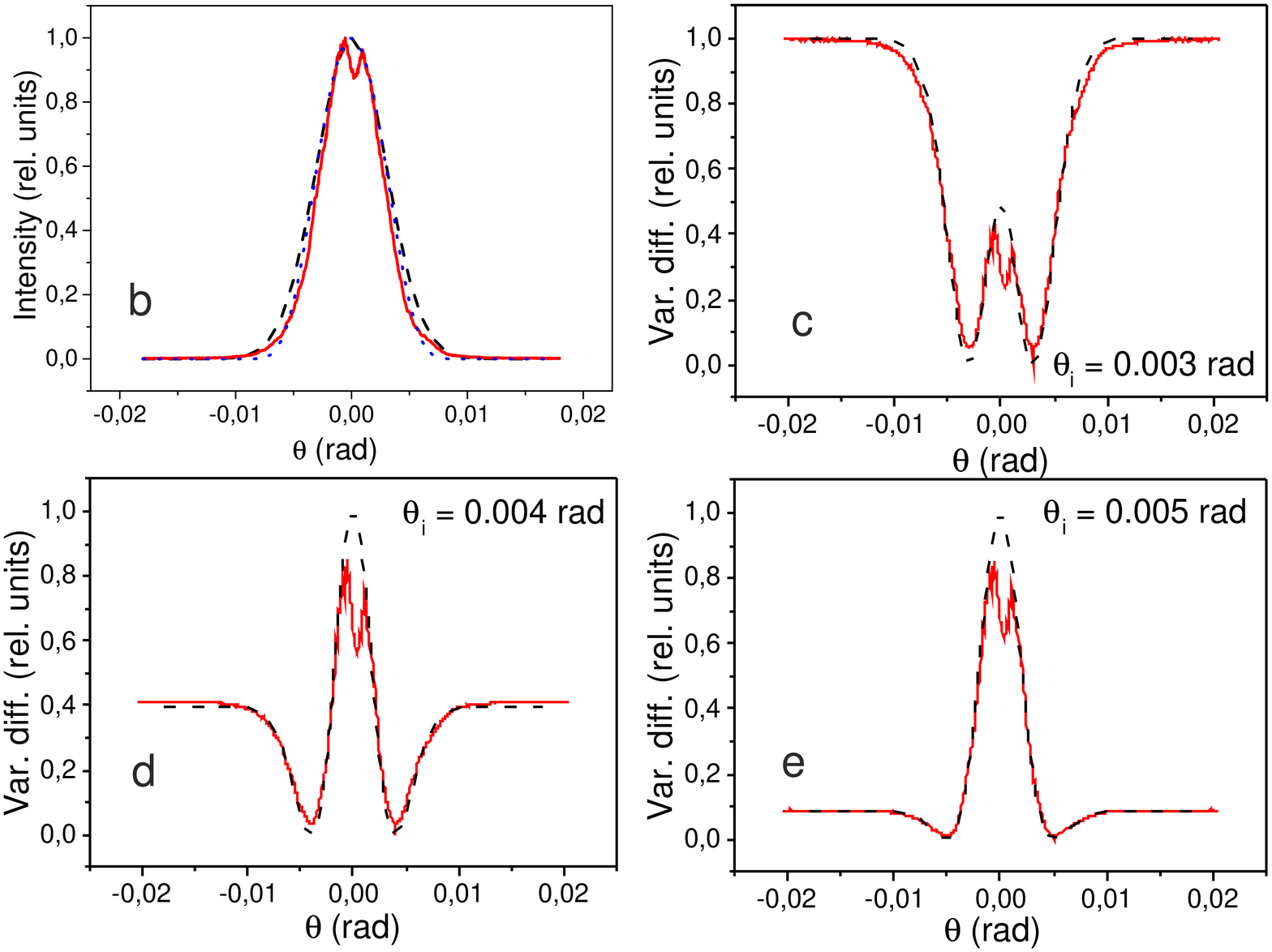}
\end{center}
\caption{The setup used for the study of BSV angular spectra (a), the measured and calculated intensity angular distributions (b) and the measured and calculated angular distributions of the photon-number difference variance with the idler angle $\theta_i$ fixed at $3$ mrad (c), $4$ mrad (d)  and $5$ mrad (e). Red solid lines: experiment, black dashed lines: theory. \label{fig:results}}
\end{figure}

\textit{Angular distributions of intensity and correlations.}
We have compared the theoretical distributions of the intensity and correlation characteristics (we used the variance of the difference intensity) with the experimental data.  BSV was generated by pumping a collinear degenerate type-I traveling-wave optical parametric amplifier (OPA) with the third harmonic of a Nd:YAG laser (wavelength $354.7$ nm, repetition rate $1$ kHz and pulse width $18$ ps) (Fig.~\ref{fig:results}a). The Gaussian pump beam was focused into a pair of $3$ mm BBO crystals by means of a lens system (L$_1$,L$_2$) that reduced the waist diameter to $120\,\mu$m FWHM. The crystals were oriented in the anisotropy-compensating configuration, that is, their optic axes were in the same plane but tilted at opposite angles to the pump~\cite{anisotropy1,anisotropy2}. They were placed at the minimum possible distance between their facets, which was $3$ mm. The expression for the TPA corresponding to this experimental setup can be found in Eq.~(1) of the Supplemental Material.  The pump was removed by means of a dichroic mirror (DM), and a red glass filter (RG645). An interference filter (IF) selected the radiation bandwidth of 10 nm around 710 nm. The 2D  spatial spectra of BSV were recorded by a CCD camera placed in the focal plane of a lens L (focal length $f=100$ mm), which, in its turn, was placed at a distance $f$ from the second crystal. The pump power used for the data collection was $20$ mW.
The effective pixel size (4.4 $\mu$m) in the Fourier plane corresponded to an angle of $0.04$ mrad, which was our resolution in the measurement of BSV angular spectra. For comparing the theoretical predictions with the measured intensity distributions and variances of the photon-number differences, we made a cut of the 2D spectrum along the axis orthogonal to the principal plane of the crystal. The profile of the spectrum along this axis (Fig.~\ref{fig:results}b) is close to Gaussian.
A reliable ensemble for calculating a statistically meaningful quantity was obtained by taking 4000 frames, during which the experimental parameters remained stable.

For all obtained angular distributions, a certain angle was fixed, for example $\theta_0$, and the difference between the CCD signals corresponding to this angle and to all other angles $\theta$ was calculated, $S(\theta)-S(\theta_0)$. The variance of this difference is a measure of intensity correlations between different plane-wave modes. Using the ensemble of frames, this variance was measured as a function of $\theta$ for several $\theta_0$. The obtained plots were normalized to the maximum of the distribution and compared with the theory. Figure \ref{fig:results} shows three cases in which different $\theta_0$ have been chosen. In red, the experimental values are presented and in black, the theoretical fits. Similarly to the measurements performed for frequency spectra in Ref.~\cite{Spasibko}, each distribution should contain, in the general case, two dips and a peak. The dips correspond to cross- and auto-correlations between the plane-wave modes while the peak, to super-bunching behavior~\cite{single-mode}.

\textit{Dependence on the parametric gain.}
According to Eq.~(\ref{new_Schmidt}), the increase in the parametric gain should lead to a re-distribution of the Schmidt eigenvalues. Low-order ones have higher values and should get enhanced, while high-order ones should get suppressed. This leads to the reduction in the effective mode number (Schmidt number),
\begin{equation}
K=\left[\sum_n\tilde{\lambda}_n^2\right]^{-1}.
\label{K}
\end{equation}
The reduction in the number of modes naturally leads to the re-shaping of the spectrum. Indeed, we measured the angular intensity distributions at different values of the parametric gain. For this experiment, we used thinner crystals (1 mm), which led to a larger mode number and a more complicated structure of the spectrum due to the interference of radiation from two separated crystals~\cite{separation}.

\begin{figure}[htb]
\begin{center}
\includegraphics[width=0.4\textwidth]{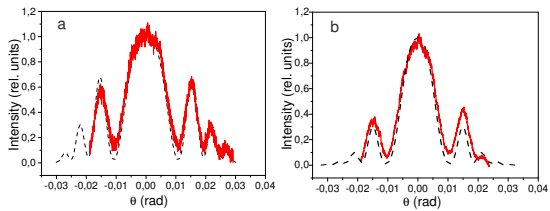}
\end{center}
\caption{Normalized angular spectra (red solid line) of PDC for two 1 mm crystals separated by 3 mm measured at the pump powers 29 mW (a) and 73 mW (b). The corresponding theoretical fits (black dashed lines) are calculated for gain values $G=2.1$ and $G=3.3$.} \label{fig:spectrum_vs_gain}
\end{figure}

\begin{figure}[htb]
\begin{center}
\includegraphics[width=0.4\textwidth]{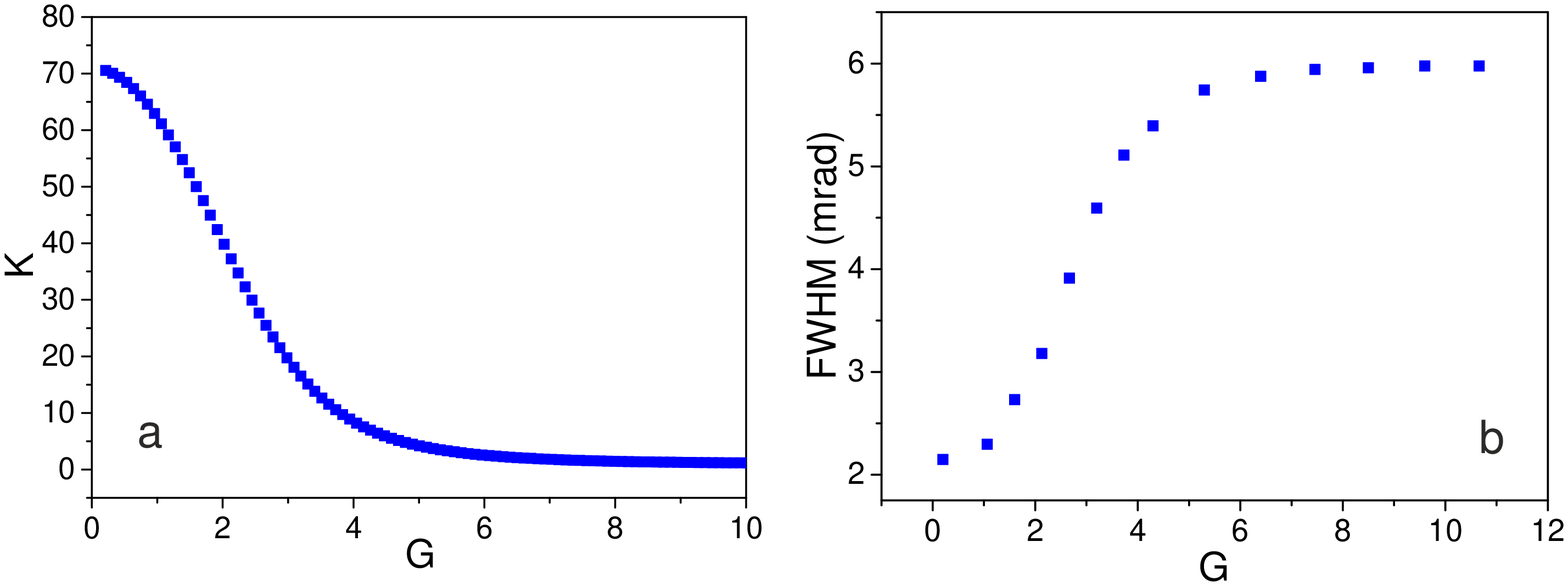}
\end{center}
\caption{The Schmidt number $K$ (a) and the width of the covariance (b) versus the parametric gain for the same experimental configuration as in Fig.~\ref{fig:spectrum_vs_gain}. }
\label{fig:modes_vs_gain}
\end{figure}

Figure~\ref{fig:spectrum_vs_gain} shows the 1D angular spectra obtained for two different pump powers, $29$ mW and $73$ mW. The positions of the minima and the maxima, given by the phase delays in the crystals and in the gap between them, do not depend on the power but the shapes of the peaks are strongly affected by it. At lower power, the side peaks are more pronounced and the central one is considerably broader than at high power. The observed shape is in agreement with the theory (black dashed lines).
Both theory and experiment show the reduction of side peaks in the spectrum, as well as the narrowing of the central peak, with the gain increase. This indirectly confirms the reduction in the number of modes. A direct calculation of the Schmidt number (\ref{K}) shows that it indeed decreases with the value of the parametric gain (Fig.~\ref{fig:modes_vs_gain}a). This is in agreement with the measurement made by Allevi and Bondani~\cite{Bondani} who observed a reduction in the Schmidt number in the range of pump powers below the depletion range. We have also calculated the angular width of the intensity covariance ~\cite{sup}, versus the parametric gain (Fig.~\ref{fig:modes_vs_gain}b). The FWHM of the covariance increases with the parametric gain, as observed in the experiments by Brida et al.~\cite{Brida2009} and Allevi et al.~\cite{Allevi}. However it appears to be limited by the size of the lowest Schmidt mode which is the only one that remains in the high-gain limit.

\begin{figure}[htb]
\begin{center}
\includegraphics[width=0.35\textwidth]{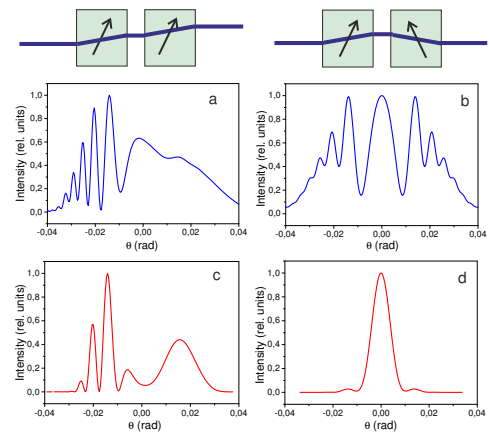}
\end{center}
\caption{Angular spectra of PDC from two crystals with the axes parallel (a,c) and tilted symmetrically with respect to the pump (b,d) calculated for the gain values $10^{-4}$ (a,b) and $10$ (c,d). On top, the crystals configurations are shown.}
\label{fig:anisotropy}
\end{figure}

\textit{Transverse walk-off and its compensation.} Finally, our model provides the description of PDC in the presence of spatial walk-off. Interestingly, manifestations of spatial walk-off at low gain and high gain are very different. While at low gain, walk-off leads only to the asymmetry of angular spectra and an increase in the number of modes~\cite{Fedorov,anisotropy2}, at high gain it can create two separate angular peaks in the geometry where collinear PDC is expected~\cite{anisotropy1}. The results of calculation are presented in Fig.~\ref{fig:anisotropy}. We consider two $1$ mm BBO crystals oriented for collinear frequency-degenerate phase matching and placed at a distance $8$ mm into a common pump beam with the FWHM waist $35\mu$m. For calculation details, see ~\cite{sup}. At low gain and the crystals optic axes parallel (Fig.~\ref{fig:anisotropy}a), spatial walk-off leads to a strong asymmetry of the spectrum while the interference of emission from different crystals creates fringes. These fringes are more pronounced on the side opposite to the walk-off direction due to the `induced coherence' effect~\cite{anisotropy1,Wang}: the radiation propagating along the pump Poynting vector (idler) from both crystals is indistinguishable, hence we see interference at matching angles (signal radiation). Placing a second crystal with the optic axis tilted symmetrically (Fig.~\ref{fig:anisotropy}b) symmetrises the spectrum but does not eliminate the fringes~\cite{anisotropy2}.

At high gain (Fig.~\ref{fig:anisotropy}c,d), the spectra of the two crystals with parallel axes in the same collinear degenerate orientation show two peaks of emission, as observed in Ref.~\cite{anisotropy1}. Our calculation reproduces these two peaks, as well as the interference fringes on one of them (Fig.~\ref{fig:anisotropy}c). The peaks are due to the fact that at high gain, amplification occurs primarily along the pump Poynting vector as well as in the direction of the twin beam. In the `compensating' configuration, at high gain no more interference fringes are observed, as seen in Fig.~\ref{fig:anisotropy}d, and strong amplification occurs in the direction collinear to the pump wave vector. The experimental distribution for a similar case is shown in Fig.~\ref{fig:results}b and is in agreement with the calculation.

\textit{In conclusion,} we have presented an analytical description of the spatial properties of BSV generated via high-gain PDC. Our theory is based on the fact that the Schmidt modes for BSV are the same as for two-photon light and only the Schmidt coefficients change at high gain. Certainly, it is not always possible to calculate analytically the modes; however, this is a different problem, common for the cases of high-gain and low-gain PDC. As long as the Schmidt modes are known, all the measurable quantities can be found analytically. The model explains several experimentally observed effects, such as the angular shape of PDC radiation and its change with the parametric gain, the dependence on the gain of the number of modes and of covariance width, the interference effects observed in a two-crystal scheme for PDC generation and, finally, the effect of spatial walk-off.

The research leading to these results has received funding from the EU FP7 under grant agreement No.
308803 (project BRISQ2). We also acknowledge partial financial support of the Russian Foundation for Basic Research, grants 14-02-31084 and 14-02-00389. We thank F.~Miatto for helpful discussions.


\begin{thebibliography}{99}
\bibitem{richer} While in low-gain PDC, correlated are only photon numbers of $0$ and $1$ and higher photon numbers occur with a very small probability, in high-gain PDC there is correlation between a larger set of photon numbers.
\end{thebibliography}
\end{document}